\documentclass[aps, prd, reprint, nofootinbib, superscriptaddress, floatfix]{revtex4-2}
\usepackage[english]{babel}
\usepackage[utf8]{inputenc}
\usepackage[T1]{fontenc}
\usepackage{hyperref, xcolor, graphicx}
\usepackage{amsfonts, amssymb, amsmath, mathtools}

\newcommand\beq{\begin{equation}}
\newcommand\eeq{\end{equation}}
\newcommand{\pd}{\partial}
\newcommand{\mO}{\mathcal{O}}

\newcommand{\tin}{\mathrm{in}}
\newcommand{\tout}{\mathrm{out}}
\newcommand{\hc}{\mathrm{H.c.}}

\hypersetup{
    colorlinks=true,
    linkcolor=black,
    citecolor=blue,   
    urlcolor=blue,
}

\begin{document}

\title{Enhancement of particle creation in nonlinear resonant cavities}

\author{Dmitrii~A.~Trunin}
    \email{dmitriy.trunin@phystech.edu}
    \affiliation{Moscow Institute of Physics and Technology, 141701, Institutskiy pereulok, 9, Dolgoprudny, Russia}

\date{\today}

\begin{abstract}
The rate of particle creation in a resonantly oscillating cavity is known to be approximately constant at large evolution times. Employing the Schwinger-Keldysh diagrammatic technique, we show that nonlinear interactions generate nonzero quantum averages and significantly enhance this rate. To illustrate this phenomenon, we consider a massless scalar field with a quartic interaction in a one-dimensional cavity with perfectly reflecting walls oscillating at twice the fundamental frequency.
\end{abstract}

\maketitle

\section{Introduction}
\label{sec:intro}

In nonstationary quantum systems, the notions of particle and vacuum state cannot be fixed once and forever, so the initial vacuum fluctuations can convert into real particles and produce measurable stress-energy fluxes ``from nothing''. This observation underlies many celebrated phenomena, including cosmological particle production~\cite{Gibbons, Birrell, Fulling:1989, Grib}, Schwinger~\cite{Grib, Schwinger:1951}, Hawking~\cite{Hawking, Hawking:1974, Hawking:1976}, and Unruh~\cite{Unruh, Fulling, Davies:1974} effects. Furthermore, the creation of real particles from vacuum fluctuations was recently observed in experiments with superconducting quantum circuits that model a resonant cavity with nonuniformly moving walls~\cite{Wilson, Svensson} or time-dependent refractive index~\cite{Lahteenmaki}. This phenomenon is known as the dynamical Casimir effect (DCE) and provides us with one of the most convenient testbeds for nonstationary quantum field theory~\cite{Nation, Dodonov:2010, Dodonov:2020}.

All these phenomena, including the DCE, are usually studied in the tree-level approximation, where interactions between the quantum fields are believed to be negligible~\cite{Birrell, Fulling:1989, Grib}. However, this approximation is generally valid only for relatively small evolution times. On the contrary, at large evolution times\footnote{The characteristic time that distinguishes these regimes is usually inversely proportional to a positive power of the coupling constant. In a closed quantum system on a stationary background, this time roughly coincides with the thermalization time~\cite{Akhmedov:review}.}, interactions build up, infrared secular memory effects become important, and loop corrections to the energy level density and correlated pair density cannot be ignored~\cite{Berges, Akhmedov:review}. In this case, the correct expressions for the created particle number and stress-energy tensor are recovered only after the resummation of the leading secularly growing loop corrections. Moreover, the resummed expressions might significantly differ from the tree-level results even if interactions are very weak. The examples of such a resummation in various nonequilibrium systems can be found in~\cite{Akhmedov:dS, Bascone, Krotov, Polyakov, Akhmedov:Et, Akhmedov:Ex, Trunin:QM}.

Recently, the DCE was also shown to suffer from the secularly growing loop corrections. In Refs.~\cite{Alexeev, Akopyan}, the secular growth was established at the two-loop level. In Ref.~\cite{Trunin:DCE}, this observation was extended to an arbitrary loop order, and the leading secularly growing corrections were resummed using a simplified quantum-mechanical Hamiltonian. Nevertheless, this approach was restricted to weak deviations from the stationarity, i.e., to such motions that produce only a small number of particles. At the same time, all experimental implementations of the DCE are based on resonant cavities, where the number and total energy of created particles rapidly grow with time~\cite{Dodonov:1993, Dodonov:1996, Lambrecht, Cole, Meplan, Li, Dalvit:1997, Dalvit:1998, Dodonov:1998, Schutzhold, Wu, Law, Crocce}. Furthermore, some of these implementations are essentially nonlinear~\cite{Lahteenmaki, Weissl, Krupko}. Hence, it is important to extend the results of~\cite{Alexeev, Akopyan, Trunin:DCE} to such nonlinear resonant cavities and check whether the interactions drastically affect the particle creation in the most feasible theoretical model of the DCE.

In this paper, we show that loop corrections significantly enhance the particle production in nonlinear resonant cavities. We consider a simple yet widespread case of resonant motion, in which the frequency of oscillations equals twice the frequency of the fundamental mode. It is remarkable that for this motion, correlation functions and quantum averages are approximately ``two-loop exact'', i.e., determined by the ``setting sun'' diagram with two classical vertices. This indicates that the late-time evolution of correlation functions and number of created particles in the resonant DCE is governed by the classical statistical approximation even if the initial quantum state is close to vacuum.

\section{Free fields}
\label{sec:free}

We begin by considering a free scalar field in a one-dimensional\footnote{We consider the one-dimensional case for two reasons. First, calculations in this case are simpler due to the absence of transverse momentum. Second, the experimental implementations of the DCE~\cite{Wilson, Svensson, Lahteenmaki} essentially work with one-dimensional scalars.} resonant cavity with perfectly reflecting walls (we assume $c = \hbar = 1$):
\beq \label{eq:free}
\left( \pd_t^2 - \pd_x^2 \right) \phi(t,x) = 0, \quad \phi\left[t,L(t)\right] = \phi\left[t, R(t)\right] = 0, \eeq
where functions $L(t)$ and $R(t)$ determine the positions of the left and right mirror, respectively. To fix the definitions of particle and vacuum in the asymptotic past, we assume that both mirrors are static, $L(t) = 0$ and $R(t) = \Lambda$, for $t < 0$. For such initial conditions, the quantized field is decomposed as follows~\cite{Moore, DeWitt, Davies:1976, Davies:1977}:
\beq \hat \phi(t,x) = \sum_{n=1}^\infty \left[ \hat a_n^\tin f_n^\tin(t,x) + \hc \right]. \eeq
Here, operators $\hat a_n^\tin$ and $\big(\hat a_n^\tin\big)^\dag$ satisfy the bosonic commutation relations, and the in-mode functions $f_n^\tin(t,x)$ are written in terms of auxiliary functions $G(z)$ and $F(z)$:
\beq \label{eq:modes}
f_n^\tin(t,x) = \frac{i}{\sqrt{4 \pi n}} \left[ e^{-i \pi n G(t+x)} - e^{-i \pi n F(t-x)} \right]. \eeq
These functions solve the generalized Moore's equations:
\beq \label{eq:Moore}
\begin{gathered}
G\left[ t + L(t) \right] - F\left[ t - L(t) \right] = 0, \\ G\left[ t + R(t) \right] - F\left[ t - R(t) \right] = 2,
\end{gathered} \eeq
and satisfy the initial conditions $G(z \le \Lambda) = F(z \le 0) = z/\Lambda$ to ensure that the in-modes have a positive definite energy in the asymptotic past:
\beq f_n^\tin(t,x) = \frac{1}{\sqrt{\pi n}} e^{-i \frac{\pi n t}{\Lambda}} \sin \frac{\pi n x}{\Lambda} \quad \text{for} \quad t < 0. \eeq
We can also define the in-vacuum as the state that is annihilated by all annihilation operators:
\beq \hat a_n^\tin | 0 \rangle = 0 \quad \text{for all} \quad n. \eeq

If the mirrors return to their initial positions and stay at rest after some interval of time $T$, we can also introduce another field decomposition:
\beq \hat \phi(t,x) = \sum_{n=1}^\infty \left[ \hat a_n^\tout f_n^\tout(t,x) + \hc \right], \eeq
where operators $\hat a_n^\tout$ and $\big(\hat a_n^\tout\big)^\dag$ again satisfy the bosonic commutation relations, and out-mode functions $f_n^\tout(t,x)$ have a positive definite energy in the asymptotic future:
\beq f_n^\tout(t,x) = \frac{1}{\sqrt{\pi n}} e^{-i \frac{\pi n t}{\Lambda}} \sin \frac{\pi n x}{\Lambda} \quad \text{for} \quad t > T. \eeq

In general, in- and out-modes do not coincide, correspond to different vacua, and are related via the canonical Bogoliubov transformation~\cite{Birrell, Fulling:1989, Grib}:
\beq \label{eq:out-to-in}
\begin{aligned}\
f_n^\tout &= \sum_k \left[ \alpha_{kn}^* f_k^\tin - \beta_{kn} \big( f_k^\tin \big)^* \right], \\
\hat a_n^\tout &= \sum_k \left[ \alpha_{kn} \hat a_k^\tin + \beta_{kn}^* \big( \hat a_k^\tin \big)^\dag \right].
\end{aligned} \eeq
The Bogoliubov coefficients:
\beq \label{eq:ab}
\alpha_{nk} = \left( f_n^\tin, f_k^\tout \right) \quad \text{and} \quad \beta_{nk} = -\left( f_n^\tin, (f_k^\tout)^* \right) \eeq
are determined by the motion of mirrors in the interval $0 < t < T$ and calculated using the Klein-Gordon inner product:
\beq \left( u, v \right) = -i \int_{L(t)}^{R(t)} \left[ u \pd_t v^* - v^* \pd_t u \right] dx. \eeq
Note that we can also consider the modes that have a positive definite energy at a fixed moment $0 < t < T$ and similarly calculate the corresponding Bogoliubov coefficients.

Now, it is straightforward to see that the expectation values of the operators $\big( \hat a_p^\tout)^\dag \hat a_q^\tout$ and $\hat a_p^\tout \hat a_q^\tout$, which have physical meaning at $t > T$, are not zero, thus indicating a difference between the in- and out-vacuum states:
\begin{widetext} \begin{gather}
\label{eq:n-out}
n_{pq}^\tout = \langle 0 | \big( \hat a_p^\tout)^\dag \hat a_q^\tout | 0 \rangle = \sum_n \beta_{np} \beta_{nq}^* + \sum_{n,k} \left[ \alpha_{np}^* \alpha_{kq} + \beta_{nq}^* \beta_{kp} \right] n_{nk}^\tin+ \sum_{n,k} \beta_{np} \alpha_{kq} \, \kappa_{nk}^\tin + \sum_{n,k} \alpha_{np}^* \beta_{kq}^* \, \kappa_{nk}^{\tin*}, \\ 
\label{eq:k-out}
\kappa_{pq}^\tout = \langle 0 | \hat a_p^\tout \hat a_q^\tout | 0 \rangle = \sum_n \alpha_{np} \beta_{nq}^* + \sum_{n,k} \left[ \beta_{np}^* \alpha_{kq} + \beta_{nq}^* \alpha_{kp} \right] n_{nk}^\tin+ \sum_{n,k} \alpha_{np} \alpha_{kq} \, \kappa_{nk}^\tin + \sum_{n,k} \beta_{np}^* \beta_{kq}^* \, \kappa_{nk}^{\tin*}.
\end{gather} \end{widetext}
We keep the initial values of $n_{nk}^\tin = \langle 0 | \big( \hat a_n^\tin)^\dag \hat a_k^\tin | 0 \rangle$ and $\kappa_{nk}^\tin = \langle 0 | \hat a_n^\tin \hat a_k^\tin | 0 \rangle$ for generality, although they are zero for a vacuum initial state. These quantum averages are referred to as the energy level density and correlated pair density, and the diagonal part $n_p = n_{pp}$ (no sum) has the meaning of the number of particles populating the $p$-th mode. Both quantities are experimentally measurable when the cavity is static and frequently used to track changes in the vacuum state (e.g., see~\cite{Lahteenmaki}). Moreover, they determine such observables as correlation functions and stress-energy fluxes.

\section{Resonant pumping}
\label{sec:resonance}

The most experimentally feasible and theoretically illustrative type of resonant motion is the one that occurs at twice the frequency of the fundamental mode:
\beq \label{eq:LR}
L(t) = 0, \quad R(t) = \Lambda \left[ 1 + \epsilon \sin\left( \frac{2 \pi t}{\Lambda} \right) \right], \eeq
where $\epsilon \ll 1$ and $0 < t < T = \Lambda \tau_f$, $\tau_f \in \mathbb{N}$, cf.~\cite{Wilson, Svensson, Lahteenmaki, Dodonov:1996}. Due to that reason, we focus on this particular type of motion in the present paper and defer its generalizations to future work. 

For resonant motion~\eqref{eq:LR}, the auxiliary functions coincide, $G(t) = F(t)$, and are given by the following approximate expression valid for times $1/\epsilon \ll t/\Lambda  \ll 1/\epsilon^2$ \cite{Dodonov:1993, Dalvit:1997, Dalvit:1998}:
\beq G(t) \approx \frac{t}{\Lambda} - \frac{1}{\pi} \arctan \frac{\left[ 1 - \zeta(t) \right] \sin\frac{2 \pi t}{\Lambda}}{\left[ 1 + \zeta(t) \right] + \left[1 - \zeta(t)\right] \cos\frac{2 \pi t}{\Lambda}} + \mO(\epsilon), \eeq
where we introduced a short notation for $\zeta(t) = e^{-2 \pi \epsilon t/\Lambda}$. Note that in the considered time interval, the function $G(t)$ rapidly approaches a staircase profile with an exponentially small width of the stair riser. For practical purposes, this profile can be approximated by a piecewise linear function:
\beq \label{eq:G-app}
G(t) \approx \begin{cases} \tau + 2 \delta \xi + \delta, \; &\text{as} \; -\frac{1}{2} \le \xi < -\delta, \\ \tau + \frac{1}{2} + \frac{1 - 2\delta + 4 \delta^2}{2\delta} \xi, \; &\text{as} \; - \delta \le \xi < \delta, \\ \tau + 1 + 2 \delta \xi - \delta, \; &\text{as} \quad \; \delta \le \xi < \frac{1}{2}. \end{cases} \eeq
Here, we parametrize the argument as $t / \Lambda = \tau + 1/2+ \xi$, $\tau \in \mathbb{N}$, $\xi \in \left[-1/2, 1/2\right)$, and approximate the half-width of the $n$-th stair riser as $\delta = \frac{1}{\pi} e^{-2 \pi \epsilon \tau}$ (so, essentially, $\tau$, $\xi$, and $\delta$ are functions of $t$).

We emphasize that the parameter $\delta$ has an important physical meaning: it determines the threshold frequency of modes that are affected by the mirror motion. Indeed, let us show that the Bogoliubov coefficients rapidly decay for mode numbers larger than $1/\delta(t)$ in the interval $1/\epsilon \ll t/\Lambda \ll 1/\epsilon^2$. Substituting the in-modes~\eqref{eq:modes} into the definitions~\eqref{eq:ab}, integrating them by parts, and employing the Moore's equations~\eqref{eq:Moore}, we get the following integral representation for the coefficients $\alpha_{nk}$ and $\beta_{nk}$:
\beq \label{eq:ab-int}
\begin{rcases} \beta_{nk} \\ \alpha_{nk} \end{rcases} = \frac{1}{2} \sqrt{\frac{k}{n}} \int_{t/\Lambda-1}^{t/\Lambda+1} e^{-i \pi n G(\Lambda z) \mp i \pi k z} dz. \eeq
Note that $|\alpha_{nk}| \approx |\alpha_{kn}|$ and $|\beta_{nk}| \approx |\beta_{kn}|$. These approximate identities are proved by integrating~\eqref{eq:ab-int} by parts and keeping in mind that the inverse function is equal to
\beq \label{eq:inverse}
G^{-1}(z)/\Lambda \approx G(\Lambda z + \Lambda/2) - 1/2 \eeq
for $\epsilon \ll 1$ and $z \gg 1/\epsilon$.

\begin{figure}
    \centering
    \includegraphics[width=\linewidth]{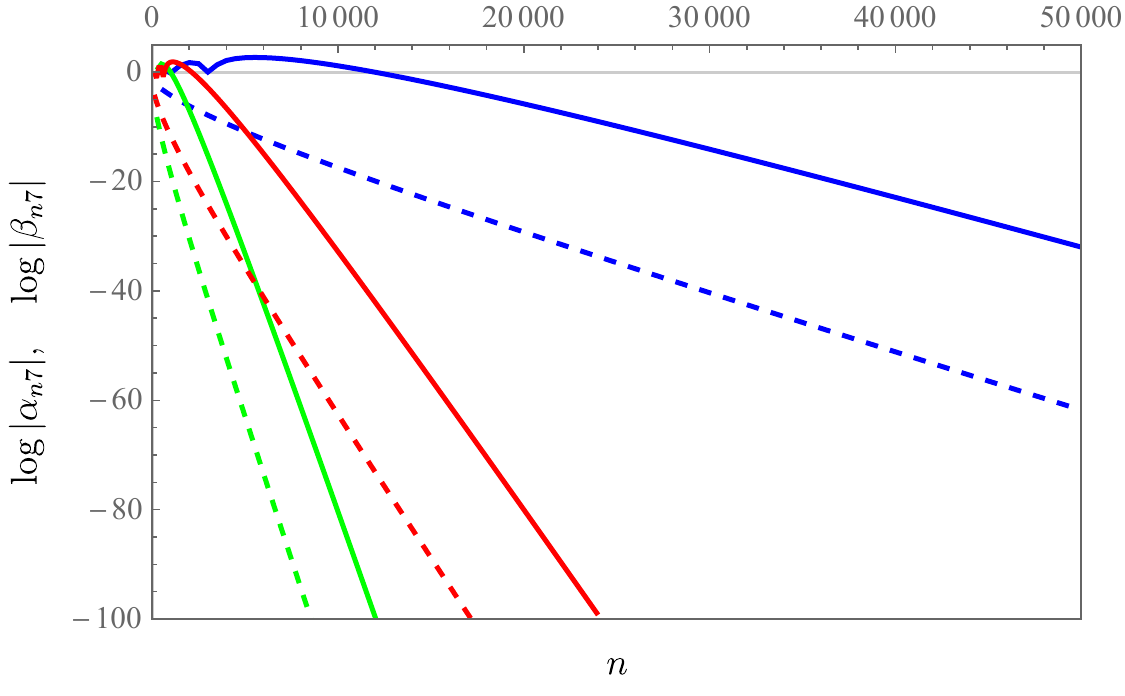}
    \caption{Numerically calculated Bogoliubov coefficients $\alpha_{n7}$ (solid lines) and $\beta_{n7}$ (dashed lines) for $\delta = 1/1000\pi$ (blue), $\delta = 1/200\pi$ (red) and $\delta = 1/100\pi$ (green).}
    \label{fig:ab-num}
\end{figure}

First, we numerically estimate integral~\eqref{eq:ab-int} for arbitrary values of $n$ and $k$, see Fig.~\ref{fig:ab-num}.
These numerical results imply that the Bogoliubov coefficients exponentially decay as
\beq \label{eq:ab-high}
\begin{rcases} \beta_{nk} \\ \alpha_{nk} \end{rcases} \sim e^{- \left( n + k \right) \delta} \quad \text{for} \quad n,k \gg 1/\delta. \eeq
So, in any expressions involving the Bogoliubov coefficients, summations over frequencies are effectively cut off at $n = 1/\delta$ and $k = 1/\delta$.

Second, we analytically calculate the integral~\eqref{eq:ab-int} for moderate frequencies employing the approximation~\eqref{eq:G-app}:
\beq \label{eq:ab-low}
\begin{rcases} \beta_{nk} \\ \alpha_{nk} \end{rcases} \approx \frac{1}{\pi} \frac{1 - (-1)^{n k}}{(-1)^{(k - 1)/2}} \frac{\sqrt{n k}}{(n \pm 2 k \delta) (k \pm 2 n \delta)}, \eeq
for $n,k \ll 1/\delta$.

Finally, keeping in mind relations~\eqref{eq:ab-high}--\eqref{eq:ab-low} and assuming $n_{pq}^\tin = 0$, $\kappa_{pq}^\tin = 0$, we estimate the energy level density and correlated pair density for a free theory~\eqref{eq:free} in the initial vacuum state:
\beq \label{eq:nk-free}
n_{pq}^\tout \approx \kappa_{pq}^\tout \approx \frac{2}{\pi} \frac{1 - (-1)^{p q}}{(-1)^{(p+q-2)/2}} \frac{1}{\sqrt{p q}} \frac{\epsilon t}{\Lambda}, \eeq
for $p,q \ll 1/\delta$ and $n_{pq}^\tout \approx \kappa_{pq}^\tout \approx 0$ otherwise. In particular, this approximate identity reproduces the rate of particle creation established in~\cite{Dodonov:1993, Dalvit:1997, Dalvit:1998, Dodonov:1996}:
\beq \label{eq:N-free}
\frac{d}{dt} n_p^\tout \approx \frac{2}{\pi} \frac{1 - (-1)^p}{p} \frac{\epsilon}{\Lambda} \quad \text{for} \quad p \ll 1/\delta, \eeq
which confirms the validity of approximations~\eqref{eq:ab-high},~\eqref{eq:ab-low}.

\section{Loop corrections}
\label{sec:loop}

Now, let us turn on interactions, i.e., consider a nonlinear generalization of the free model~\eqref{eq:free}:
\beq \left( \pd_t^2 - \pd_x^2 \right) \phi(t,x) = -\frac{\pd V(\phi)}{\pd \phi}, \eeq
where $V(\phi) = \sum_{a=3}^\infty \lambda_a \phi^a$. To be specific, we first discuss the quartic interaction:
\beq \label{eq:quartic}
\left( \pd_t^2 - \pd_x^2 \right) \phi(t,x) = -\lambda \phi^3(t,x), \eeq
and then generalize the results to arbitrary potentials.

In nonstationary situations, interactions lead to two separate effects~\cite{Akhmedov:review}. On one hand, they affect the particle production during the period of nonstationarity ($0 < t < T$ in the model~\eqref{eq:LR}), i.e., generate non-trivial loop corrections to the initial quantum averages $n_{pq}^\tin$ and $\kappa_{pq}^\tin$. On the other hand, interactions force the system to thermalize in the asymptotically static regions (at least, in dimensions higher than two). In this paper, we are interested in the particle production during the mirror's motion, so we consider the initial vacuum state and assume that the coupling constant $\lambda$ is adiabatically turned on after $t = t_0 < 0$ and abruptly turned off after $t = T$. In other words, we estimate the corrections to $n_{pq}^\tout$ and $\kappa_{pq}^\tout$ in the interacting theory by the moment $t = T$. The further evolution of these quantum averages, which, we believe, describes the thermalization of the system, will be studied elsewhere.

In general, loop corrections to $n_{pq}^\tin$ and $\kappa_{pq}^\tin$ are conveniently calculated in the Schwinger-Keldysh diagram technique~\cite{Schwinger, Keldysh, Kamenev, Rammer, Landau:vol10, Arseev}. In the particular model~\eqref{eq:quartic}, this technique contains two interaction vertices:
\beq \label{eq:vertices}
-i \lambda \int_{t_0}^T dt \int_{L(t)}^{R(t)} \!\!dx \, \phi_{cl}^3 \phi_q, \quad -i \frac{\lambda}{4} \int_{t_0}^T dt \int_{L(t)}^{R(t)} \!\!dx \, \phi_{cl} \phi_q^3, \eeq
and three propagators:
\beq \begin{aligned}
G_{12}^\mathrm{K(eldysh)} &= -i \big\langle \phi_{cl}(t_1, x_1) \phi_{cl}(t_2, x_2) \big\rangle, \\
G_{12}^\mathrm{R(etarded)} &= -i \big\langle \phi_{cl}(t_1, x_1) \phi_q(t_2, x_2) \big\rangle, \\
G_{12}^\mathrm{A(dvanced)} &= -i \big\langle \phi_q(t_1, x_1) \phi_{cl}(t_2, x_2) \big\rangle,
\end{aligned} \eeq
where angle brackets denote the averaging w.r.t. some initial state, and $\phi_{cl}$ and $\phi_q$ correspond to the classical and quantum components of the field following the notations of the Schwinger-Keldysh technique. At the tree level, retarded and advanced propagators characterize the particle spectrum and do not depend on the initial state:
\beq \label{eq:free-R}
i G_{12}^\mathrm{R,free} = i G_{21}^\mathrm{A,free} = \theta(t_1 - t_2) \sum_n \left( f_{1,n}^\tin f_{2,n}^{\tin*} - \hc \right), \eeq
where we introduce the short notation $f_{a,n}^\tin = f_n^\tin(t_a, x_a)$. On the contrary, the tree-level Keldysh propagator is determined by initial quantum averages of interest to us:
\beq \label{eq:free-K}
i G_{12}^\mathrm{K,free} \!=\! \sum_{p,q} \!\left[ \!\left(\frac{\delta_{pq}}{2} + n_{pq}^\tin \right)\! f_{1,p}^\tin f_{2,q}^{\tin*} + \kappa_{pq}^\tin f_{1,p}^\tin f_{2,q}^\tin + \hc \right]\!. \eeq
Furthermore, propagators approximately preserve the form~\eqref{eq:free-R},~\eqref{eq:free-K} if both their points are taken to the future infinity while the difference is kept finite~\cite{Akhmedov:review, Berges, Kamenev}:
\beq \label{eq:limit}
\frac{t_1 + t_2}{2} \gg \frac{1}{\lambda \Lambda} \quad \text{and} \quad \frac{t_1 + t_2}{2} \gg |t_1 - t_2|. \eeq
This property allows us to extract the corrected quantum averages in the interacting theory from the exact Keldysh propagator in the limit in question.

To estimate the loop resummed Keldysh propagator in the interacting model~\eqref{eq:quartic} with resonantly moving mirrors~\eqref{eq:LR} in the limit~\eqref{eq:limit}, we make four crucial observations.

First, we map the trajectories~\eqref{eq:LR} to stationary ones:
\beq t + x = G^{-1}(\tau + \xi), \quad t - x = G^{-1}(\tau - \xi), \eeq
in all internal vertices of diagrams that describe loop corrections to the Keldysh propagator, e.g.:
\beq \begin{aligned}
V &= -i \lambda \int_{t_0}^T dt \int_{L(t)}^{R(t)} \!\!dx \, f_m^\tin(t,x) f_n^\tin(t,x) f_p^\tin(t,x) f_q^\tin(t,x) \\
&= -i \lambda \int_{\tau_0}^{\tau_f} d\tau \int_0^1 d\xi \, \frac{d G^{-1}(\tau - \xi)}{d \tau} \frac{d G^{-1}(\tau + \xi)}{d \tau} \times \\ &\times e^{-i \pi (m+n+p+q)\tau} \frac{\sin(\pi m \xi) \sin(\pi n \xi) \sin(\pi p \xi) \sin(\pi q \xi)}{\pi^2 \sqrt{m n p q}}.
\end{aligned} \eeq

Second, we expect that the leading contribution to the loop corrections come from large evolution times: $t/\Lambda = \tau \gg 1/\epsilon$. Keeping in mind identities~\eqref{eq:G-app} and~\eqref{eq:inverse}, we approximate the $G^{-1}$ with a piecewise-linear function:
\beq \label{eq:V2}
\begin{aligned}
V &\approx -i \lambda \Lambda^2 \int_{1/\epsilon}^{\tau_f} d\tau \int_0^1 d\xi \, \delta_\delta(\tau - \xi) \delta_\delta(\tau + \xi) \times \\ &\times e^{-i \pi (m+n+p+q)\tau} \frac{\sin(\pi m \xi) \sin(\pi n \xi) \sin(\pi p \xi) \sin(\pi q \xi)}{\pi^2 \sqrt{m n p q}}.
\end{aligned} \eeq
Here, $\delta_\delta(\tau) = 1/2\delta_s$ if $\big| \tau - s - 1/2 \big| < \delta_s$ for some $s \in \mathbb{N}$, $\delta_\delta(\tau) = 2 \delta_s$ otherwise, and $\delta_s = \frac{1}{\pi} e^{-2 \pi \epsilon s}$.

Third, the exponential decay of the Bogoliubov coefficients~\eqref{eq:ab-high} and the relation~\eqref{eq:out-to-in} between the in-modes and the positive energy modes at a moment $t$ imply that sums over the virtual momenta are effectively cut off\footnote{More precisely, sums over larger momenta give negligible corrections to~\eqref{eq:nk-loop} if we consider times $1/\lambda \Lambda^2 \ll t/\Lambda \ll 1/(\lambda \Lambda^2)^2$.} at mode numbers $n \sim 1/\delta_s$. Roughly speaking, we can exclude high-energy modes from consideration because they are unaffected by the mirror motion and are not involved in the particle creation\footnote{However, we cannot use $1/\delta_s$ as an ultraviolet regulator for other purposes, e.g., renormalization of mass and coupling constant.}. Keeping in mind this cutoff, we replace the functions $\delta_\delta(\tau)$ with exact delta-functions and reduce integrals~\eqref{eq:V2} to sums:
\beq V \approx -i \lambda \Lambda^2 \sum_{s=1/\epsilon}^{\tau_f} g_m^s g_n^s g_p^s g_q^s, \eeq
where we introduce the notation for the ``remnant'' of the initial mode $f_n^\tin(t,x)$:
\beq \label{eq:g}
f_n^\tin(t,x) \to g_n^s = -i \frac{(-1)^s}{\sqrt{\pi n}} \frac{1 - (-1)^n}{2}. \eeq

\begin{figure}
    \centering
    \includegraphics[width=\linewidth]{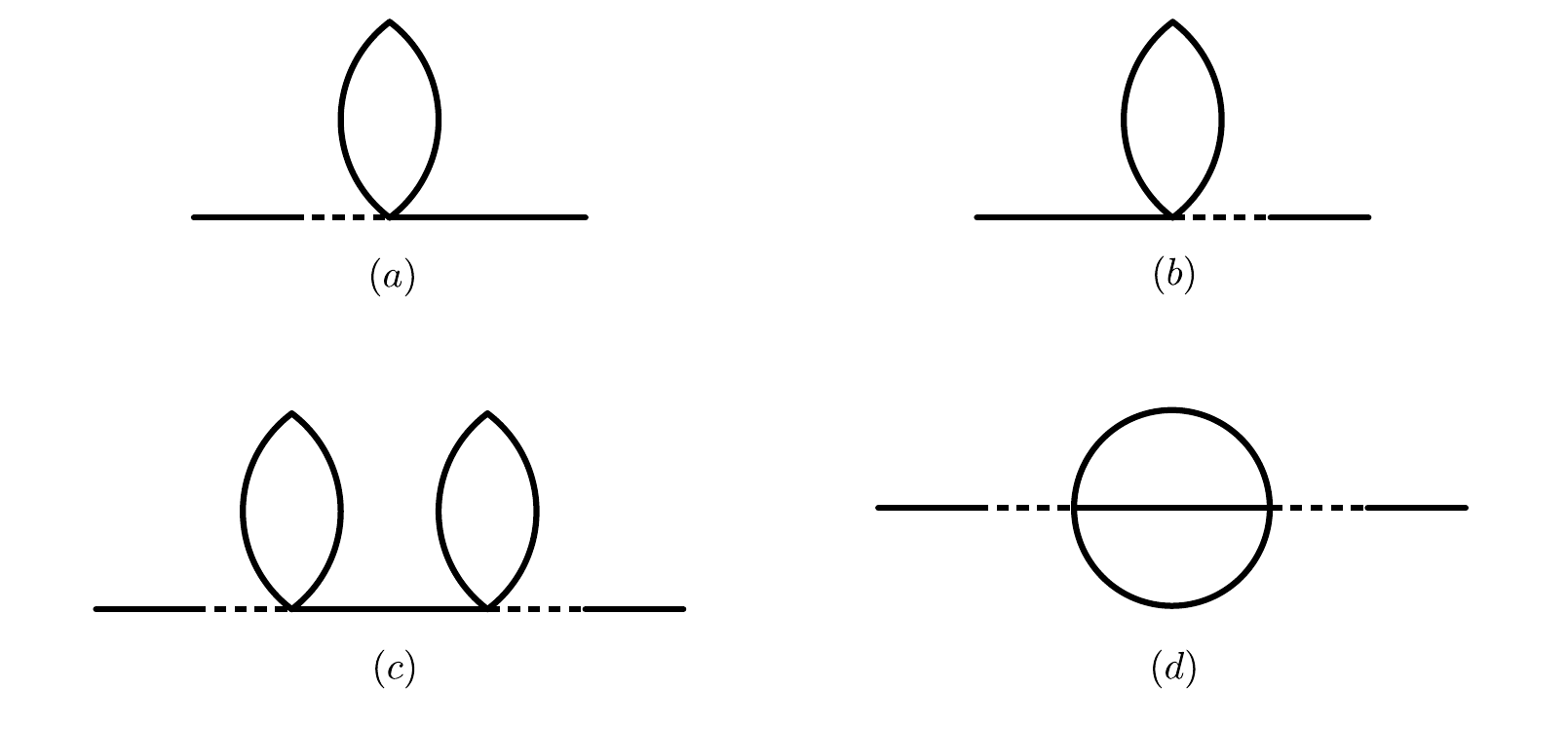}
    \caption{Loop corrections to the Keldysh propagator that do not contain internal retarded/advanced propagators. Solid lines denote the tree-level Keldysh propagators, half-dashed lines denote the retarded/advanced propagators.}
    \label{fig:loops}
\end{figure}

Fourth, now, it is straightforward to see that any diagrams containing virtual retarded/advanced propagators are approximately zero. On one hand, if retarded propagator~\eqref{eq:free-R} connects two internal vertices, both $f_{1,n}^\tin$ and $f_{2,n}^\tin$ are eventually replaced with their ``remnants''~\eqref{eq:g}. On the other hand, functions $g_n^s$ are purely imaginary, and their product is purely real, so the r.h.s. of~\eqref{eq:free-R} is approximately zero\footnote{This reasoning is based on the behavior of $G(z)$, so it does not apply directly to higher-order resonances or nonresonant motions.}. Therefore, as long as we are interested only in the leading contribution to the exact Keldysh propagator in the interacting theory, we can consider only such loop diagrams where internal vertices are connected by the Keldysh propagators alone (see Fig.~\ref{fig:loops}).

Furthermore, the ``tadpole'' diagrams (Fig.~\ref{fig:loops}a--c) can be absorbed into the renormalized mass and do not contribute to the evolution of the initial state and particle creation~\cite{Alexeev, Akopyan, Trunin:DCE}. Hence, we need to estimate only the ``setting sun'' diagram (Fig.~\ref{fig:loops}d). Keeping in mind approximations~\eqref{eq:V2}--\eqref{eq:g} and employing decomposition~\eqref{eq:free-K}, we obtain the leading correction to $n_{pq}^\tin$ and $\kappa_{pq}^\tin$:
\beq \label{eq:nk-loop}
\Delta n_{pq}^\tin \approx \Delta \kappa_{pq}^\tin \approx \frac{3}{5 \pi} \frac{1 - (-1)^{pq}}{2 \sqrt{p q}} \big( \lambda \Lambda T \big)^2 \left( \frac{\epsilon T}{\Lambda} \right)^3, \eeq
for $p,q \ll 1/\delta_{\tau_f} \sim e^{2 \pi \epsilon T/\Lambda}$.

Finally, we substitute $\Delta n_{pq}^\tin$ and $\Delta \kappa_{pq}^\tin$ into Eqs.~\eqref{eq:n-out}--\eqref{eq:k-out} and take into account Eq.~\eqref{eq:nk-free} to determine the relative correction to the quantum averages $n_{pq}^\tout$ and $\kappa_{pq}^\tout$, which are physically meaningful in the asymptotic future:
\beq \label{eq:ratio}
\frac{\Delta n_{pq}^\tout}{n_{pq}^\tout} \approx \frac{\Delta \kappa_{pq}^\tout}{\kappa_{pq}^\tout} \approx \frac{12}{5} \big( \lambda \Lambda T \big)^2 \left( \frac{\epsilon T}{\Lambda} \right)^4. \eeq
So, loop contributions to the energy level density and correlated pair density significantly exceed the tree-level expressions in the time interval $1/\lambda \Lambda^2 \ll T/\Lambda \ll 1/(\lambda \Lambda^2)^2$ and $1/\epsilon \ll T/\Lambda \ll 1/\epsilon^2$. 

The ``two-loop exactness'' of quantum averages~\eqref{eq:nk-loop} and~\eqref{eq:ratio} resembles the ``one-loop exactness'' of scalar electrodynamics on a strong electric field background~\cite{Akhmedov:Et, Akhmedov:Ex}. In that case, $n_{pq}$ and $\kappa_{pq}$ are also determined by the first nontrivial loop correction and uniformly grow at large evolution times due to the symmetry of mode functions. 

We also emphasize that the argumentation of this section is equally applicable to general analytic potentials\footnote{We exclude the powers of $\phi$ less than third because $V(\phi) \sim \phi^0$ and $V(\phi) \sim \phi^1$ do not affect the dynamics, and we assume that the field is massless, so $\lambda_2 = m^2/2 = 0$.}, $V(\phi) = \sum_{a=3}^\infty \lambda_a \phi^a$. Repeating it for each power of $\phi$, we obtain an analog of Eq.~\eqref{eq:ratio}:
\beq \label{eq:ratio-gen}
\frac{\Delta n_{pq}^\tout}{n_{pq}^\tout} \approx \frac{\Delta \kappa_{pq}^\tout}{\kappa_{pq}^\tout} \approx \sum_{a=2}^\infty \frac{8 \, (2a)!}{2a+1} (\lambda_{2a} \Lambda T)^2 \left( \frac{\epsilon T}{\Lambda} \right)^{2a}, \eeq
where we took into account that the odd powers are suppressed in the limit $1/\lambda \Lambda^2 \ll T/\Lambda \ll 1/(\lambda \Lambda^2)^2$ and $1/\epsilon \ll T/\Lambda \ll 1/\epsilon^2$. In particular, for the experimentally and theoretically interesting potential
\beq \label{eq:sin-gordon}
V(\phi) = \lambda \left[ \cos(\phi) + \phi^2/2 - 1 \right], \eeq
we get $\lambda_{2a} = \lambda (-1)^a/(2a)!$ and
\beq \Delta n_{pq}^\tout \approx \Delta \kappa_{pq}^\tout \sim (\lambda \Lambda T)^2 \left[ \sinh \left(\frac{\epsilon T}{\Lambda} \right) - \frac{\epsilon T}{\Lambda} \right]. \eeq
Thus, for interactions~\eqref{eq:sin-gordon}, number of particles created \textit{in each mode} grows exponentially instead of linearly.

\section{Discussion}
\label{sec:discussion}

We have shown that the energy level density and the correlated pair density, which are generated during resonant oscillations of the cavity walls in an interacting theory, are significantly enhanced compared to a free theory. We emphasize that the enhancement factor~\eqref{eq:ratio} or~\eqref{eq:ratio-gen} is determined by the coupling constants $\lambda_a$ and relative amplitude of wall oscillations $\epsilon$, but does not depend on the mode numbers $p$ and $q$ for the low-laying modes. Hence, the number of particles created in each mode, the total number of particles, and their total energy are enhanced by the same factor~\eqref{eq:ratio}. 

In particular, such an enhancement implies that the number of particles created in even modes is small in both linear and nonlinear models, as would be expected for resonant pumping~\cite{Dodonov:1993, Dalvit:1997, Dalvit:1998, Dodonov:1996}. At the same time, note that the number of particles created in separate modes grows slower than exponential (unless the interaction term is specially chosen), which distinguishes the resonant DCE from other parametric resonances.

Note that the leading loop corrections to the correlation functions and quantum averages~\eqref{eq:ratio}--\eqref{eq:ratio-gen} are given by the ``setting sun'' diagrams with two classical vertices\footnote{That is, vertices with only one quantum component, e.g., the first vertex in~\eqref{eq:vertices}.}. Such a behavior indicates that the non-perturbative dynamics may be well described by the classical statistical approximation, so higher-order corrections can be calculated within the semiclassical approach~\cite{Berges, Berges:2007, Aarts:2001, Aarts:1997, Radovskaya}. However, we emphasize that classical statistical approximation is valid only when all modes are highly populated. In the DCE with the initial vacuum state, this condition is indeed fulfilled at large evolution times, but violated at the very beginning of evolution. This observation distinguishes the DCE from the standard applications of the semiclassical approach.

Our results can be extended in several possible directions. First, the enhancement~\eqref{eq:ratio}--\eqref{eq:ratio-gen} encourages a careful measurement of the large-time asymptotics of $n_{pq}^\tout$ and $\kappa_{pq}^\tout$ in experimental implementations of the DCE~\cite{Wilson, Svensson, Lahteenmaki, Nation} (see also Refs.~\cite{Dodonov:2022, Recamier, Sousa, Srivastava} for other nonlinear models of the DCE). We emphasize that this enhancement is noticeable only at evolution times $T/\Lambda \gg 1/\lambda \Lambda^2$. Although such evolution times are by two orders larger than the typical measurement time in existing experiments, we expect them to be achieved in the future modifications~\cite{Trunin:DCE}.

Second, the nonlinear DCE is very similar to an interacting scalar field in a rapidly expanding universe~\cite{Moschella, Pavlenko, Anempodistov, Bazarov, Serreau-1, Serreau-2} or its condensed matter analogs~\cite{Jain, Carusotto, Zache, Chatrchyan, Butera, Barroso, Eckel}. On the one hand, loop corrections to
the Keldysh propagator grow equally rapidly with time for the DCE and for the light scalar field in an expanding Poincar\'{e} patch of de~Sitter space. On the other hand, condensed matter analogs of particle creation in an expanding universe are closely related to parametric instabilities and can be described using a semiclassical approach. So, it is interesting to study the calculations of Sec.~\ref{sec:loop} in light of this relation.

Finally, it is promising to extend the results of this paper to resonant oscillations different from~\eqref{eq:LR} and study the evolution of quantum averages~\eqref{eq:nk-loop} in the asymptotic future $t > T$.

\begin{acknowledgments}
We are grateful to Emil Akhmedov and Damir Sadekov for valuable discussions and proofreading of the paper. We also thank an anonymous referee for helpful comments. This work was supported by the Russian Ministry of education and science and by the grant from the Foundation for the Advancement of Theoretical Physics and Mathematics ``BASIS''.
\end{acknowledgments}

\end{document}